\documentclass[twocolumn,footinbib,floatfix,aps,pra,superscriptaddress,notitlepage]{revtex4-1}
\usepackage[usenames,dvipsnames]{xcolor} 
\usepackage[normalem]{ulem}

\usepackage{amsmath,amssymb}
\usepackage{amsfonts}
\usepackage{bm}
\usepackage{bbm}
\usepackage[dvipsnames]{xcolor}
\usepackage{latexsym}
\usepackage[english]{babel}
\usepackage{times}
\usepackage{stmaryrd}
\usepackage{psfrag,graphicx}
\usepackage{appendix}
\usepackage{enumitem}
\usepackage{bbold}
\usepackage{xr}
\usepackage[colorlinks=true,citecolor=red,linkcolor=brown,urlcolor=orange]{hyperref}

\newcommand{\be}{\begin{equation}}
\newcommand{\ee}{\end{equation}}
\newcommand{\ben}{\begin{eqnarray}}
\newcommand{\een}{\end{eqnarray}}

\begin{document}

\title{Bounding the quantum limits of precision for phase estimation with loss and thermal noise}

\author{Christos N. Gagatsos}
\affiliation{Department of Physics, University of Warwick, Coventry CV4 7AL, UK}
\author{Boulat A. Bash}
\affiliation{Quantum Information Processing Group, Raytheon BBN Technologies, Cambridge, Massachusetts, USA 02138}
\author{Saikat Guha}
\affiliation{Quantum Information Processing Group, Raytheon BBN Technologies, Cambridge, Massachusetts, USA 02138}
\affiliation{College of Optical Sciences, University of Arizona, 1630 East University Boulevard, Tucson, AZ 85719, USA}
\author{Animesh Datta}
\affiliation{Department of Physics, University of Warwick, Coventry CV4 7AL, UK}

\date{\today}

\begin{abstract}
We consider the problem of estimating an unknown but constant carrier phase modulation $\theta$ using a general -- possibly entangled -- $n$-mode optical probe through $n$ independent and identical uses of a lossy bosonic channel with additive thermal noise. We find an upper bound to the quantum Fisher information (QFI) of estimating $\theta$ as a function of $n$, the mean and variance of the total number of photons $N_{\rm S}$ in the $n$-mode probe, the transmissivity $\eta$ and mean thermal photon number per mode ${\bar n}_{\rm B}$ of the bosonic channel. Since the inverse of QFI provides a lower bound to the mean-squared error (MSE) of an unbiased estimator $\tilde{\theta}$ of $\theta$, our upper bound to the QFI provides a lower bound to the MSE. It already has found use in proving fundamental limits of covert sensing, and could find other applications requiring bounding the fundamental limits of sensing an unknown parameter embedded in a correlated field.
\end{abstract}

\maketitle

\section{Introduction}
Loss and noise are inevitable in all physical systems. Indeed, lossy bosonic channels with additive thermal noise henceforth called lossy thermal-noise channels are ubiquitous -- in communications across fibres and free-space links as well as wireless sensor networks. Although the number of thermal photons at optical wavelengths is small at room temperature, amplification in an optical channel provides an effective environment of thermal noise. The quantum communication limits of optical channels in terms of channel capacity have long been studied \cite{Holevo1997,Caves1994,GGL2004,Konig2013,Giovannetti2014}. As quantum sensing moves into practical applications, the quantum limits of sensing capabilities in optical channels will become increasingly relevant. However, there have been few studies on the quantum limits of sensing capabilities in optical channels \cite{Takeoka2016,Lupo2017}. In particular, as the lossy thermal-noise channel is an accurate quantum description of many optical channels, estimation of unknown carrier phase modulation $\theta$ over this channel is a problem of wide and imperative appeal in quantum sensing. Our work addresses it.

\begin{figure}
\includegraphics[scale=0.5]{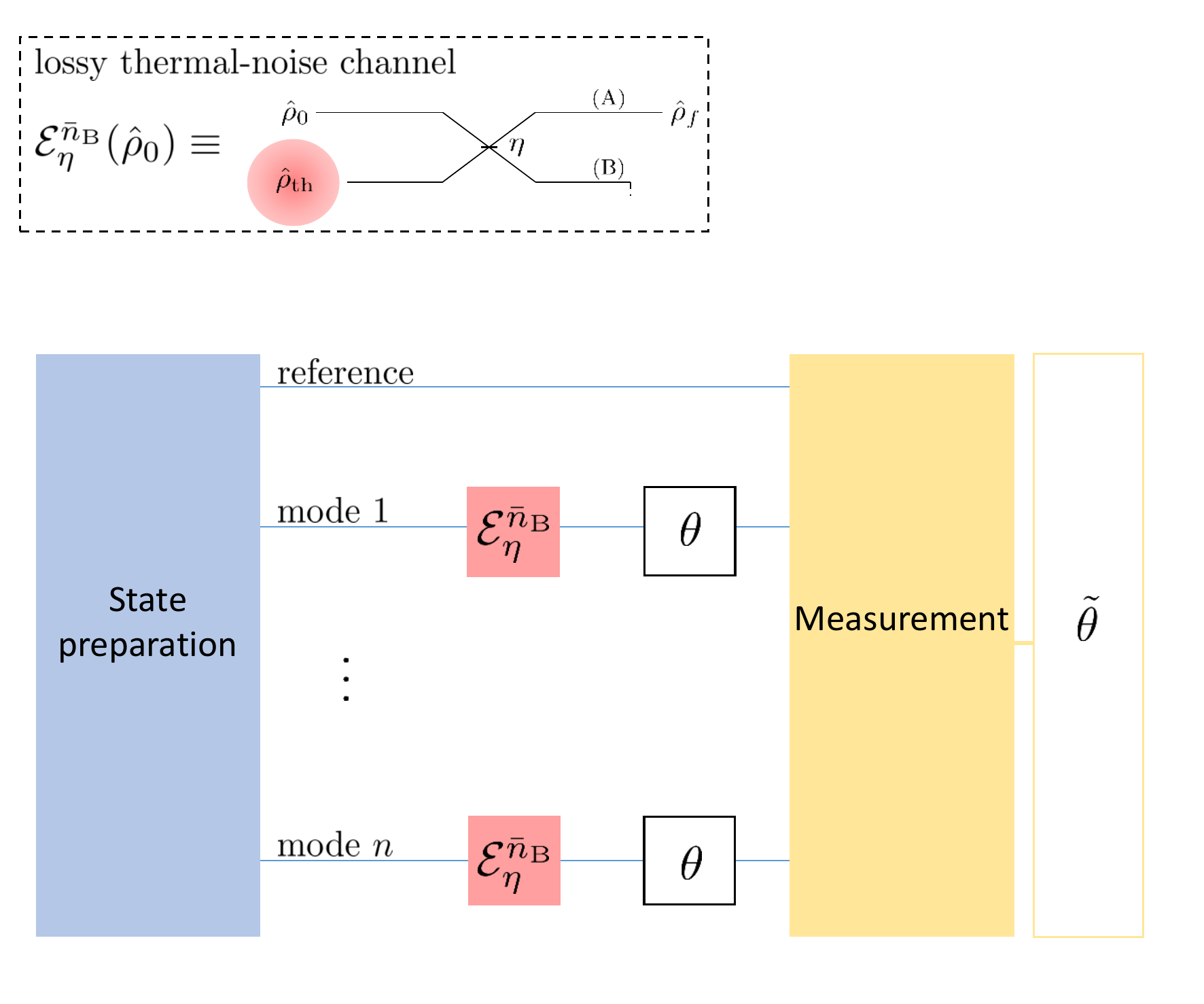}
\caption{Phase estimation over a lossy thermal-noise channel. We consider the use of a general $n$-mode optical probe for estimating an unknown but constant phase modulation $\theta$, where each mode of the probe is independently corrupted by an identical lossy thermal-noise channel $\mathcal{E}^{\bar{n}_{\rm B}}_{\eta}$ with mean thermal noise photon number $\bar{n}_{\rm B}$ and transmissivity $\eta$.  We are interested in a lower bound on the mean squared error $\langle (\theta-\tilde{\theta})^2\rangle$ of the phase estimator $\tilde{\theta}$.}
\label{setup}
\end{figure}

We consider the problem of estimating unknown but constant carrier phase modulation $\theta$ using an $n$-mode optical probe through $n$ independent and identical uses of lossy thermal-noise channel, as described in Fig.~\ref{setup}. A fully general probe may be entangled across modes. The performance of this sensing task can be captured by the variance of an unbiased estimator $\tilde{\theta}$ and is limited by the quantum Cram\'er-Rao bound
\begin{eqnarray}
\label{eq:qcrb}
\langle (\theta - \tilde{\theta})^2 \rangle \geq F_\mathrm{Q}^{-1}(\theta),
\end{eqnarray}
where $F_\mathrm{Q}(\theta)$ is the quantum Fisher information (QFI) associated with phase $\theta$, and may be a function of $\theta$ ~\cite{Helstrom:1976aa,PARIS:2009aa}.

Analytically closed expressions for the QFI are in general difficult to obtain. Several works \cite{Giov2006, Paris2009, Helstrom1976} provide a fundamental and attainable bound, however, it cannot be calculated analytically. Therefore, it has little value to understanding the problem at hand nor can be applied in specific tasks such as covert sensing \cite{Bash2017}. However, one can extract many features of precision of estimators through analytical upper bounds on the QFI~\cite{Fujiwara2008,Escher2011,Escher2012,Demkowicz2012,Demkowicz2013,Jarzyna2017}. In this work we provide such bound for QFI associated with phase $\theta$, denoted by $C_\mathrm{Q}(\theta)$. We \emph{lower-bound the lower bound} on the variance of the estimator $\tilde{\theta}$ by combining Eq.~\eqref{eq:qcrb} with the relation
\begin{eqnarray}
\label{Bound0}F_\mathrm{Q}(\theta) &\leq & C_\mathrm{Q}(\theta) \Rightarrow F_\mathrm{Q}^{-1}(\theta) \geq C_\mathrm{Q}^{-1}(\theta).
\end{eqnarray}
Our main result, Eq.~(\ref{FinalBound_main}), is, to the best of our knowledge, the first upper bound on the QFI that accounts for thermal photons in the environment and an arbitrary input state.
Our bound is a function of the number of modes $n$, the mean and variance of the total number of photons $N_{\rm S}$ in the $n$-mode probe, and the channel transmissivity $\eta$ and mean thermal photon number per mode ${\bar n}_{\rm B}$.
Obtaining Eq.~\eqref{FinalBound_main} requires non-trivial optimisation to ensure that it is tight and decreasing with increasing thermal noise.
It has already found use in proving the fundamental limits of covert sensing \cite{Bash2017}. 
Other potential applications of our result include finite-length analysis of channel estimation in quantum key distribution protocols \cite{Leverrier2015}, distributed sensing using shared entanglement \cite{Proctor2017,Ge2017,Zhuang2017,Peter2013,Magda2016}, and other problems requiring bounding the fundamental limits of sensing an unknown parameter embedded in a correlated field.

We denote operators with a circumflex (e.g., $\hat{\rho}$ for density operators) and estimators with a tilde (e.g., $\tilde{\mu}$ or $\tilde{\theta})$. Bold capital letters stand for operators (such as $\mathbf{K}$), which we use when we refer to Kraus operators as well.

\section{Bounding the quantum Fisher information}
Consider the problem of estimating a parameter $\mu$ from a state $\hat{\rho}_{\mu}$, which can be an output of a channel characterized by $\mu$. Denote the QFI for the parameter $\mu$ estimated from the state $\hat{\rho}_{\mu}$ by $F_\mathrm{Q}(\mu)$. 
While closed-form expressions for QFI for a single phase have been found for pure states \cite{PARIS:2009aa} and general Gaussian states \cite{Monras2013,Pirandola2015}, no such formulae are known for arbitrary quantum states. Numerical methods can provide the exact QFI in specific systems and scenarios, but closed-form expressions are by definition more powerful and, thus, desirable. Finally, they are valuable for optimising the performance of a given sensing set-up over arbitrary probe states and essential in proving optimality in general.

Let $C_\mathrm{Q}(\mu)$ be the QFI for the parameter $\mu$ estimated from a \emph{purification} $|\Psi(\mu)\rangle$ of the state $\hat{\rho}_{\mu}$.
Since, we can extract more information about the parameter when the system and the environment are monitored together rather than monitoring the system alone~\cite{Escher2011} $F_\mathrm{Q}(\mu) \leq C_\mathrm{Q}(\mu)$.
If the evolution of the system from some initial state to $\hat{\rho}_{\mu}$ is described by the set of Kraus operators $\{\mathbf{K}_{\mathbf{l}}(\mu)\}$, where $\mathbf{l}$ may refer to multiple 
indices (non-bold $l$ always refers to single index), it has been shown \cite{Escher2011} that
\begin{eqnarray}
\label{upperDef}C_\mathrm{Q}(\mu) &=& 4 \left[\langle\hat{H}_1 \rangle - \langle\hat{H}_2 \rangle^2 \right],\\
\label{H1Def}\hat{H}_1 &=& \sum_{\mathbf{l}}  \frac{d\mathbf{K}_\mathbf{l}^\dagger}{d\mu} \frac{d\mathbf{K}_\mathbf{l}}{d\mu},\\
\label{H2Def}\hat{H}_2 &=& i \sum_{\mathbf{l}}  \frac{d\mathbf{K}_\mathbf{l}^\dagger}{d\mu} \mathbf{K}_\mathbf{l},
\end{eqnarray} 
where the mean values in Eq.~(\ref{upperDef}) are taken on the input state, which can be pure or mixed. For brevity we have suppressed the dependence of $\mathbf{K}_{\mathbf{l}}$ on $\mu$.
The bound in Eq.~(\ref{upperDef}) can be generalised to the case of $n$ identical channels \cite{Escher2011} to
\begin{eqnarray}
C_{Q,n}(\mu) &=& 4 \sum_{m_0=1}^n \big(\langle \hat{H}_1^{(m_0)} \rangle - \langle \hat{H}_2^{(m_0)} \rangle^2  \big)\\
\nonumber && + 8 \sum_{m_1 = 2}^{n} \sum_{m_2=1}^{m_1-1} \big(\langle \hat{H}_2^{(m_1)} \hat{H}_2^{(m_2)} \rangle  -  \langle \hat{H}_2^{(m_1)} \rangle \langle \hat{H}_2^{(m_2)} \rangle  \big),
\label{multibound}
\end{eqnarray}
where the mean values are taken over the input $n$-mode state, while $H_{1,2}^{(m_i)}$ refers to the standard definitions of Eqs.~(\ref{H1Def}), (\ref{H2Def}) for the $m_i$-th quantum channel.

The purification of a quantum state is not unique. Therefore, in seeking the tightest bound, which is $C_\mathrm{Q}(\mu)=F_\mathrm{Q}(\mu)$ \cite{Escher2011}, we must optimise over all possible purifications or at least selectively optimise over some possible purifications. The non-uniqueness of the purification is linked to the unitary ambiguity of Kraus operators, since both of these ambiguities are rooted in the freedom of choosing the environments' basis up to some local unitary. Specifically, the unitary ambiguity of Kraus operators means the following: two Kraus representations $\{\mathbf{K}_n\}$ and $\{\mathbf{K'}_n\}$ represent the same quantum channel if and only if  $\mathbf{K}_n = \sum_{n,m} U_{mn} \mathbf{K'}_m$, where $U_{mn}$ are the elements of the matrix representation of a unitary operator that acts on the environment's Hilbert space. 
Optimisation over all possible equivalent representations of a quantum channel is in general formidable. However, a limited optimisation over a subset of equivalent Kraus representations should yield better results than no optimisation. 
In particular, the aim is to minimise the amount of information about the parameter $\mu$  in the environment. 
The more of this information is erased by the local unitary operations, the tighter is the inequality (\ref{Bound0}).

\section{Lossy thermal-noise channel and phase shift}
In the lossy thermal-noise channel the input state $\hat{\rho}_0$ interacts with a thermal state $\hat{\rho}_{\textrm{th}}$ via a beam splitter of transmissivity $\eta$. Then the environment is traced out, leaving the channel's output state $\hat{\rho}_{f}$.
The transformation $\hat{U}$ for the action of the beam splitter on a single mode is
\begin{eqnarray}
\hat{U} = \begin{pmatrix}
\sqrt{\eta} & \sqrt{1-\eta} \\
-\sqrt{1-\eta} & \sqrt{\eta}
\end{pmatrix},
\end{eqnarray}
and the thermal state can be expressed in Fock basis as
\begin{eqnarray}
\hat{\rho}_{\textrm{th}} = \frac{1}{1+\bar{n}_\mathrm{B}} \sum_{k=0}^{\infty} \left( \frac{\bar{n}_\mathrm{B}}{1+\bar{n}_\mathrm{B}} \right)^k | k\rangle \langle k |,
\end{eqnarray} where $\bar{n}_\mathrm{B}$  is the mean thermal photon number.

To proceed, we need the Kraus operators for the lossy thermal-noise channel. One possible Kraus operator description is found by decomposing the lossy thermal-noise channel into a pure loss channel of transmissivity $\tau=\eta/G$ followed by a quantum-limited amplifier with gain $G=1+(1-\eta) \bar{n}_\mathrm{B}$ \cite{Ivan2011}.
This provides a Kraus representation for the lossy thermal-noise channel as $\hat{\rho}_{f} = \sum_{k,l=0}^{\infty} \mathbf{B}_k \mathbf{A}_l \hat{\rho}_0 \mathbf{A}_l^\dagger \mathbf{B}_k^\dagger$,
where $\mathbf{A}_l$ are the Kraus operators of the pure loss channel, while $\mathbf{B}_k(G)$ are the Kraus operators of the quantum limited amplifier \cite{Ivan2011} given by
\begin{eqnarray}
\label{PureLoss1}\mathbf{A}_l &=& \sqrt{\frac{(1-\tau)^l}{l!}} \tau^{\frac{\hat{n}}{2}} \hat{a}^l,\\
\label{ThermalNoise1}\mathbf{B}_k &=& \sqrt{ \frac{1}{k!} \frac{1}{G} \left(\frac{G-1}{G}\right)^k} \hat{a}^{\dagger k} G^{-\frac{\hat{n}}{2}}. 
\end{eqnarray}

The parameter to be estimated is the phase picked up by the input state, that is $\mu \equiv \theta$. The Kraus representation of the full channel including the phase shift $\theta$ is
\be
\hat{\rho} = \sum_{k,l=0}^{\infty} e^{i \theta \left[ \hat{n} +\gamma (k-l) \right]} \mathbf{B}_k \mathbf{A}_l \hat{\rho}_0 \mathbf{A}_l^\dagger \mathbf{B}_k^\dagger e^{-i \theta \left[ \hat{n} +\gamma (k-l) \right]},\phantom{1}
\label{kraus} 
\ee
where, as it will become apparent, $\gamma\ \in \mathbb{R}$ controls the position of the phase shift operator with respect to the lossy thermal-noise channel.
 
The state in Eq.~(\ref{kraus}) is independent of $\gamma$. Therefore, the position of the phase operator with respect to the lossy thermal-noise channel does not affect the QFI $F_\mathrm{Q}(\theta)$~\cite{mankei2013}. However, 
the position of the phase shift impacts the bound $C_\mathrm{Q}(\theta)$, as can be see from Eqs.~(\ref{upperDef}), (\ref{H1Def}), and (\ref{H2Def}). The case for which $\gamma=-1$ corresponds to the phase shift being applied 
before the lossy thermal-noise channel, while the case $\gamma=0$ corresponds to the phase shift being applied after it. This follows from the identity $e^{i \theta \left(\hat{n}-(k-l)\right)} \hat{a}^{\dagger k} G^{-\frac{\hat{n}}
{2}} \tau^{\frac{\hat{n}}{2}} \hat{a}^l = 
\hat{a}^{\dagger k} G^{-\frac{\hat{n}}{2}} \tau^{\frac{\hat{n}}{2}} \hat{a}^l e^{i \theta \hat{n}}$, which can be proven using the relation $\left(\hat{a}^\dagger\right)^l \hat{n}^k= \left(\hat{n} - l \right)^k \left(\hat{a}^\dagger
\right)^l.$ This latter relation is also useful for the calculations that follow.

Our strategy for extracting a tighter bound is to optimise over the Kraus operators $\mathbf{A}$ and $\mathbf{B}$ by adding local phase shift operators, whose generators depend linearly on the estimated parameter (see App. Sec. \ref{PhysMean}). We impose the dependence of the Kraus operators $\mathbf{A}$ and $\mathbf{B}$ on $\theta$ so that we obtain non-trivial results when using Eqs.~(\ref{H1Def}) and (\ref{H2Def}). These local phases generate equivalent Kraus decompositions of the lossy thermal-noise channel. Hence this is a special case of the local unitary freedom on the environment in defining the Kraus operator.
To that end, consider two equivalent Kraus representations of the pure loss channel, i.e., $\{\mathbf{A}_l\}$ of Eq.~(\ref{PureLoss1}) and $\{\mathbf{A'}_l\}$, and let the unitary operator connecting them be a local phase rotation by $x \theta$. That is,
\begin{eqnarray}
\nonumber \mathbf{A'}_l &=& \sum_{k=0}^{\infty} \delta_{kl} e^{i x k \theta} \mathbf{A}_l = e^{i x l \theta} \mathbf{A}_l\\
&=& \sqrt{\frac{(1-\tau)^l}{l!}} \tau^{\frac{\hat{n}}{2}} \left(e^{i x \theta}\hat{a}\right)^l,
\label{PureLoss2}
\end{eqnarray}
where $x\ \in \mathbb{R}$. Therefore, the operators in Eq.~(\ref{PureLoss2}) can be obtained from Eq.~(\ref{PureLoss1}) by applying a local phase $\theta x$ in the system's modes, $e^{-i \theta x \hat{n}} \hat{a} e^{i \theta x \hat{n}} = e^{i \theta x} \hat{a}$. 
Thus, we define a family of Kraus representations, parametrised by $x$, which we optimise over the continuous real parameter $x$. Following the same reasoning, we can define a family of Kraus representations $\{\mathbf{B'}_k\}$ for quantum-limited amplification channel as
\begin{eqnarray}
\nonumber \mathbf{B'}_k &=& e^{i \theta y k}\mathbf{B}_k\\
&=&\sqrt{ \frac{1}{k!} \frac{1}{G} \left(\frac{G-1}{G}\right)^k} \left( e^{i \theta y} \hat{a}^{\dagger}\right)^k G^{-\frac{\hat{n}}{2}},
\label{QuantAmpl2}
\end{eqnarray}   
where now the family of equivalent Kraus operators is parametrised by the continuous parameter $y\ \in \mathbb{R}$.
Now we can define a Kraus representation for the full channel, i.e., phase shift, loss and thermal noise,
\begin{eqnarray}
\nonumber \hat{\rho} &=& \sum_{k,l=0}^{\infty}  e^{i \theta \hat{n}}  \mathbf{B}_k e^{i \theta k y} \mathbf{A}_l e^{i \theta l x} \hat{\rho}_0 \mathbf{A}_l^\dagger e^{-i \theta l x} \mathbf{B}_k^\dagger e^{-i \theta k y} e^{-i \theta \hat{n}}\\
&=& \sum_{k,l=0}^{\infty} e^{i \theta \left( \hat{n} +l x + k y \right)} \mathbf{B}_k \mathbf{A}_l \hat{\rho}_0 \mathbf{A}_l^\dagger \mathbf{B}_k^\dagger e^{-i \theta \left( \hat{n} + l x +k y \right)},
\label{krausXY} 
\end{eqnarray}
where $x$ and $y$ do not affect the evolution of the initial state and, therefore, the value of $F_\mathrm{Q}(\theta)$ (nor of any other quantity that depends only in the properties of the channel's output state), but they have an impact on $C_\mathrm{Q}(\theta)$. Therefore, we optimise $C_\mathrm{Q}(\theta)$ over $x$ and $y$. Note Eq.~(\ref{kraus}) is a special case of Eq.~(\ref{krausXY}) for $-x=y=\gamma$. This gives the physical meaning for considering $\gamma\ \in \mathbb{R}$: it is a local phase shift in the environment's degrees of freedom.

\section{Bound for single-mode lossy thermal-noise channel}
In order to compute the upper bound $C_\mathrm{Q}(\theta)$ on the QFI corresponding to the output state of the lossy thermal-noise channel, for estimating the phase $\theta$, we apply 
Eqs.~(\ref{upperDef}), (\ref{H1Def}), and (\ref{H2Def}) for the Kraus operator $\mathbf{K_l}=e^{i \theta \left( \hat{n} +l x + k y \right)} \mathbf{B}_k \mathbf{A}_l,$ where $\mathbf{l}=l,k$,
\begin{eqnarray}
\label{H1}\hat{H}_1 &=& \sum_{k,l=0}^{\infty}  \mathbf{A}_l^\dagger \mathbf{B}_k^\dagger \frac{d e^{-i \theta \left( \hat{n} +l x + k y \right)} }{d\theta} \frac{d e^{i \theta \left( \hat{n} +l x + k y \right)}}{d\theta} 
\mathbf{B}_k \mathbf{A}_l,\phantom{12}\ \\
\label{H2}\hat{H}_2 &=& i \sum_{k,l=0}^{\infty}  \mathbf{A}_l^\dagger \mathbf{B}_k^\dagger \frac{d e^{-i \theta \left( \hat{n} +l x + k y \right)}}{d\theta} e^{i \theta \left( \hat{n} +l x + k y \right)} \mathbf{B}_k 
\mathbf{A}_l.\phantom{12}\ 
\end{eqnarray}
We find (see App. Sec. \ref{CalcsSingle}) that the bound minimised over $x,y$ is 
\begin{eqnarray}
\label{SingleModeBound}C_\mathrm{Q}^\star(\theta) = \frac{4 \eta \langle \Delta n_S^2 \rangle \bar{n}_S [\eta \bar{n}_S + (1-\eta)\bar{n}_\mathrm{B} +1]}{D},
\end{eqnarray}
where
\begin{eqnarray}
\nonumber D&=&(1-\eta)\langle \Delta n_\mathrm{S}^2 \rangle [\eta \bar{n}_\mathrm{S} (2 \bar{n}_\mathrm{B}+1) - \eta \bar{n}_\mathrm{B} (\bar{n}_\mathrm{B}+1)\\
&& + (\bar{n}_\mathrm{B}+1)^2] + \eta \bar{n}_\mathrm{S} [\eta \bar{n}_\mathrm{S} + (1-\eta) \bar{n}_\mathrm{B} + 1]
\end{eqnarray}
and $\langle \Delta n_\mathrm{S}^2 \rangle = 
\langle \hat{n}^2 \rangle - \langle \hat{n} \rangle^2 = \langle \hat{n}^2 \rangle - \bar{n}_\mathrm{S}^2$ is the variance of the system's photon number. It can be shown that $C_\mathrm{Q}^\star(\theta)$
is a decreasing function of the environment's mean thermal photon number $\bar{n}_\mathrm{B}$ (see App. Sec. \ref{CalcsSingle}) and therefore the bound obtained behaves 
reasonably, meaning that the QFI decreases when the environmental temperature is increased.

\section{Bound for $n~$independent and identical lossy thermal-noise channel}
We consider $n$ identical lossy thermal-noise channel with the same mean thermal photons number $\bar{n}_\mathrm{B}$, equivalently we consider $n$
uses of the same lossy thermal-noise channel. For that case we find the bound (see App. Sec. \ref{App2}),
\begin{eqnarray}
\nonumber C_{\mathrm{Q},n}^\star(\theta) &=& \frac{1}{D_n}\Big\{ 4 n \eta \langle \Delta N_\mathrm{S}^2 \rangle \langle N_\mathrm{S} \rangle [1+\bar{n}_\mathrm{B} (1-\eta)]\\
&& + 4 \eta^2 \langle \Delta N_\mathrm{S}^2 \rangle \langle N_\mathrm{S} \rangle^2\Big\},
\label{FinalBound_main}
\end{eqnarray}
where the denominator reads,
\begin{eqnarray}
\nonumber D_n&=&\eta^2 \langle N_\mathrm{S} \rangle^2 + \eta n \langle N_\mathrm{S} \rangle [1+(1-\eta)\bar{n}_\mathrm{B}]\\
\nonumber &&+ (1-\eta)\eta \langle \Delta N_\mathrm{S}^2 \rangle \langle N_\mathrm{S} \rangle (1+2 \bar{n}_\mathrm{B})\\
\nonumber &&- (1-\eta) \eta \langle \Delta N_\mathrm{S}^2 \rangle n \bar{n}_\mathrm{B} (1+\bar{n}_\mathrm{B})\\
&&+(1-\eta)n \langle \Delta N_\mathrm{S}^2 \rangle (1+\bar{n}_\mathrm{B})^2
\end{eqnarray}
and $\langle N_\mathrm{S} \rangle$ and $\langle \Delta N_\mathrm{S}^2 \rangle$ are respectively the mean photon and photon number variance of the $n$-mode input state. 
It can be shown that the bound $C_{\mathrm{Q},n}^\star(\theta)$ is a decreasing function of the environment's mean thermal photon number (see App. Sec. \ref{App2}).

\begin{figure}
\includegraphics[scale=0.5]{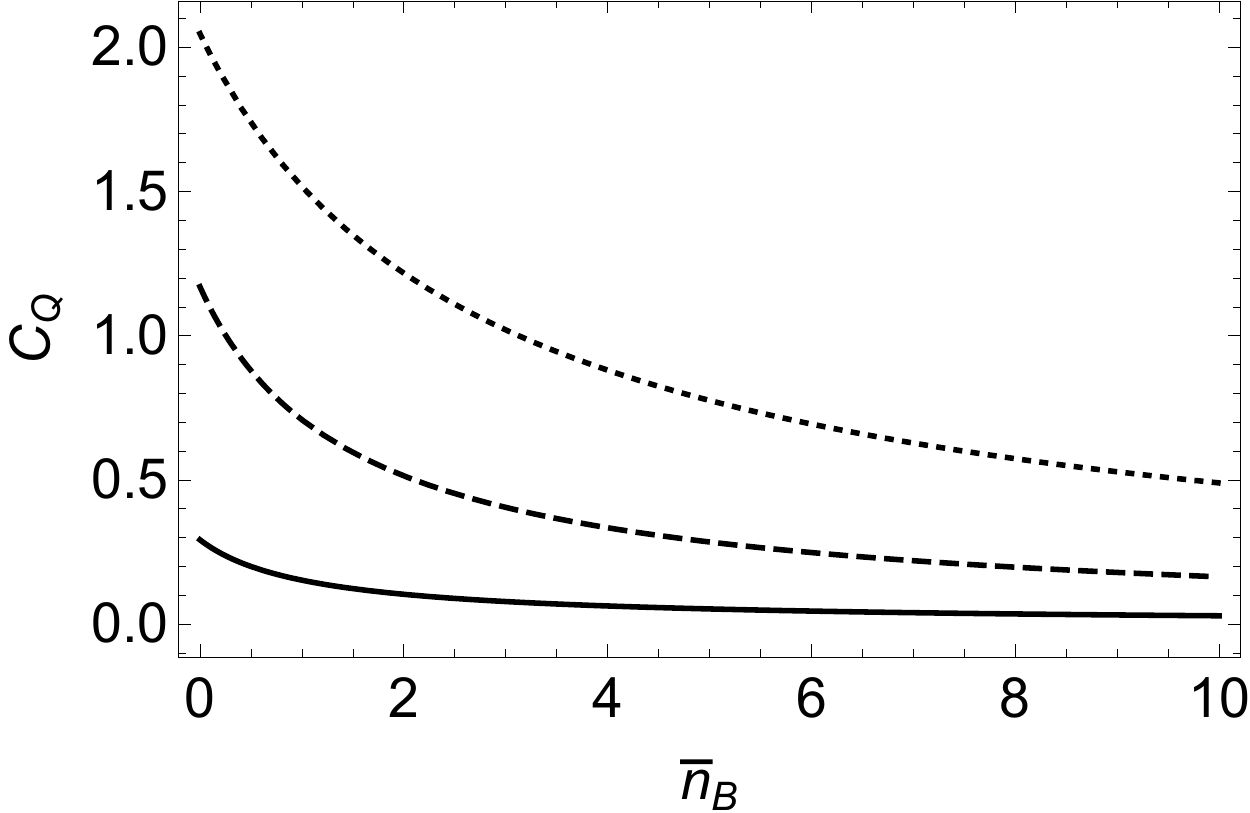}
\caption{The bound corresponding to an entangled coherent probe state as a function of mean thermal photon number. The amplitude is $|\alpha|=1$ for all curves. Solid curve corresponds to $\eta=0.1$, dashed curved corresponds to $\eta=0.4$, and dotted curved corresponds to $\eta=0.7$.}
\label{plot}
\end{figure}

As an example, we consider an entangled coherent state \cite{Sanders2012} of the form,
\begin{eqnarray}
|\Psi\rangle = \frac{1}{\sqrt{2(1+e^{-|\alpha|^2})}} \left(|\alpha,0\rangle + |0, \alpha \rangle\right)
\label{ecs}
\end{eqnarray}
where $|\alpha \rangle$ is a coherent state. Both modes of the entangled coherent state suffer the same losses and thermal noise and pick up the same phase shift. The mean photon number $\langle N_\mathrm{S} \rangle$ and the photon number variance $\langle \Delta N_\mathrm{S}^2 \rangle$ for the state (\ref{ecs}) are
\begin{eqnarray}
\langle N_\mathrm{S} \rangle &=& \frac{|\alpha|^2}{1+e^{-|\alpha|^2}}\\
\langle \Delta N_\mathrm{S}^2 \rangle &=&  \frac{|\alpha|^2}{1+e^{-|\alpha|^2}}.
\end{eqnarray} 
The bound can then be found from Eq. (\ref{FinalBound_main}) and we plot its behaviour with respect to mean thermal photon number in Fig. \ref{plot}.
 
\section{Conclusions}
We have analytically bounded the QFI associated with phase modulation of an arbitrary probe state (single-mode and $n$-mode) that suffers thermal noise and loss. The bound we have derived in Eq.~(\ref{FinalBound_main}) behaves reasonably with thermal photons, i.e., it decreases with $\bar{n}_\mathrm{B}$. We note that our bound is sub-optimal, as better bounds may be found by choosing different unitarily equivalent Kraus decompositions of the lossy thermal-noise channel. However, the task of further optimisation is onerous since one can only hope for an educated guess on the local unitary transformation which acts on the environment. Based on the physical argument of erasing the phase information leaked to the environment, we have chosen the phase shift operator as the local unitary to act on the environment. Beyond this, one is left with a trial and error approach.

Moreover, we show that the position of the phase shift with respect to the lossy thermal-noise channel aids optimisation of the bound but is irrelevant in the calculation of the Fisher information to be bounded. Note that this proof is valid for a larger class of quantities, i.e., we have shown that for every quantity $\mathcal{Q}$ that depends only on the properties of the final reduced density matrix, the position of the phase shift operator with respect to the lossy thermal-noise channel does not play any role in the calculation of $\mathcal{Q}$.

Since the lossy thermal-noise channel is an accurate quantum description of many practical communication channels, having a worked out bound on the output's QFI for sensing phase modulation through such channel is important and useful. Our bound has already found use in covert sensing \cite{Bash2017}, where thermal noise is inevitable and the main ingredient at the same time.  Therefore, we anticipate that our results are applicable in a wide range of problems requiring bounding the fundamental limits of sensing an unknown parameter embedded in a correlated field.

\begin{acknowledgments}
CNG and AD acknowledge the UK EPSRC (EP/K04057X/2) and the UK National Quantum Technologies Programme (EP/M01326X/1, EP/M013243/1).  BAB and SG acknowledge the support from Raytheon BBN Technologies, DARPA under contract number HR0011-16-C-0111 and ONR under prime contract number N00014-16-C-2069.
\end{acknowledgments}

\bibliography{BoundTLCBib}

\newpage

\onecolumngrid
\appendix
\section*{Appendix}
\renewcommand{\thesubsection}{\arabic{subsection}}

\subsection{Physical meaning of the equivalent Kraus operators}
\label{PhysMean}
\begin{figure}[h]
\includegraphics[scale=0.5]{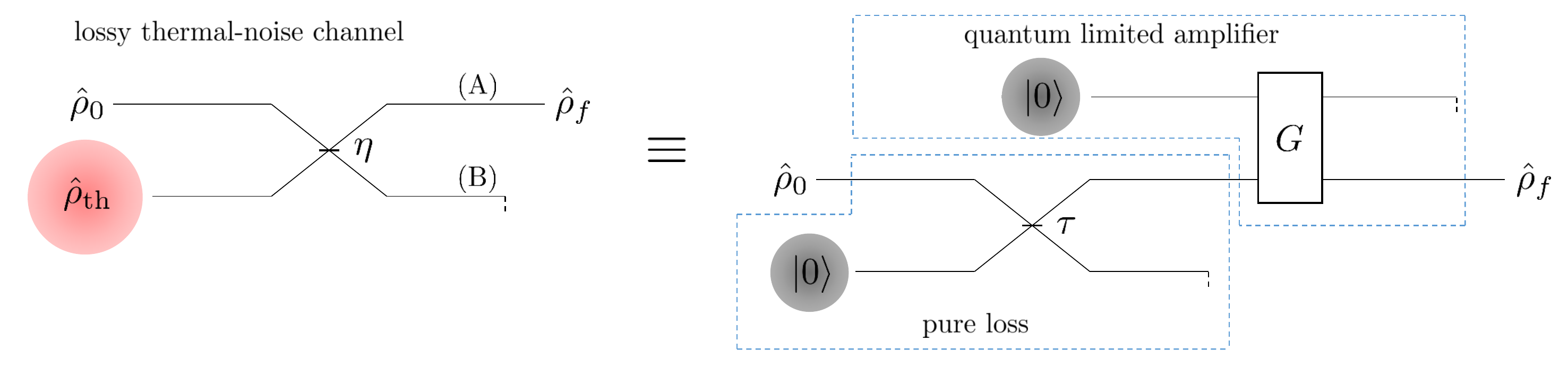}
\caption{The lossy thermal-noise channel (left) can be decomposed into a pure loss channel followed by a quantum limited amplifier (right). If the lossy thermal-noise channel is characterised by transmissivity $\eta$ and mean thermal photon number $\bar{n}_\mathrm{B}$, then the pure loss channel is characterised by transmissivity $\tau=\eta/G$ and the quantum limited amplifier has gain $G=1+(1-\eta) \bar{n}_\mathrm{B}$.}
\label{FigThermalLosses}
\end{figure}

\begin{figure}
\includegraphics[scale=0.5]{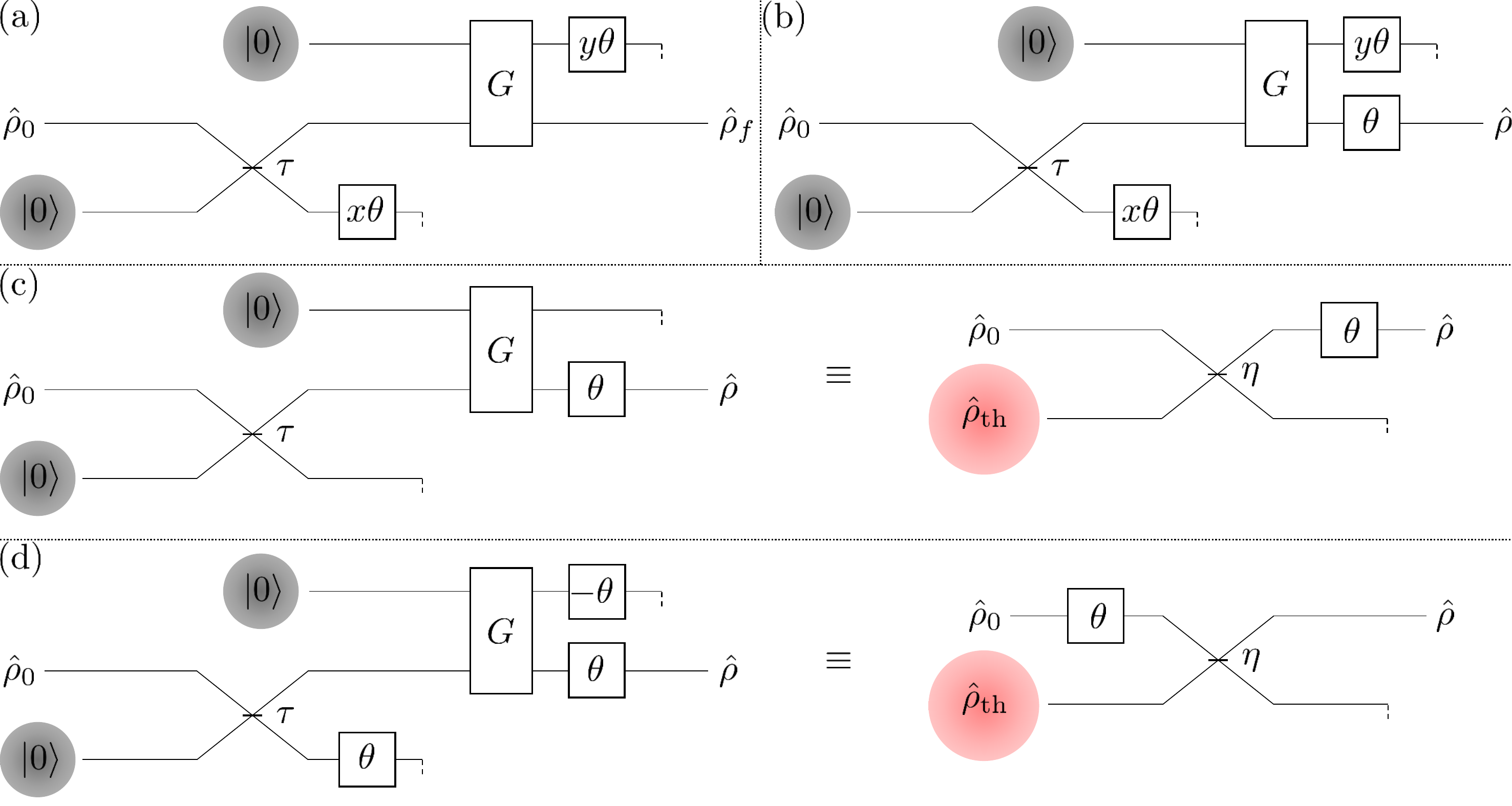}
\caption{(a) A Kraus decomposition of the lossy thermal-noise channel equivalent to that depicted in Fig.~\ref{FigThermalLosses}. The local phases applied to the environment have no effect on $F_\mathrm{Q}(\theta)$ but they affect the bound $C_\mathrm{Q}(\theta)$.
(b) The initial state suffers from loss and additive thermal noise, and then acquires a phase $\theta$.
(c) For $x=y=0$ the loss and thermal noise precedes the phase shift.
(d) For $x=-y=1$ the phase shift precedes the loss and thermal noise.
We minimise $C_\mathrm{Q}(\theta)$ over $x,y \in \mathbb{R}$, i.e., over the family of equivalent Kraus decompositions.}
\label{FigThermalLossesPhases}
\end{figure}

\def\theequation{A\arabic{equation}}
\setcounter{equation}{0}
We prove that the unitary equivalence between the sets of Kraus operators we have chosen in the main text has the physical interpretation of a local phase in the environment after interaction.
First of all we decompose the lossy thermal-noise channel into a pure loss channel followed by  quantum-limited amplifier \cite{Ivan2011}, as depicted in Fig. \ref{FigThermalLosses}.
Let $\hat{U}$ be a beam splitter transformation with transmissivity $\tau$, the Kraus operators for the pure loss channel is defined as
\be
\label{PureLossDef}\mathbf{A}_l = \langle l | \hat{U} | 0 \rangle = \sum_{m,n=0}^{\infty} \langle m,l | \hat{U} | n,0 \rangle |m\rangle \langle n|,
\ee
where $m,n$ refer to the system while $l$ and the vacuum state $|0\rangle$ refer to the environment. Also, note that in Eq.~(\ref{PureLossDef}) we use the Fock basis for the system and the environment. Employing the unit resolution of the coherent basis in Eq.~(\ref{PureLossDef}) we obtain
\begin{eqnarray}
\nonumber \mathbf{A}_l &=& \frac{1}{\pi^2} \sum_{m,n=0}^{\infty} \int d^2 \alpha_1 \int d^2 \alpha_2  \exp\Bigg( -\frac{|\alpha_1|^2}{2} -\frac{|\alpha_2|^2}{2} -\frac{1}{2} |\alpha_1 \sqrt{\tau}+\alpha_2 \sqrt{1-\tau} |^2 - \frac{1}{2} |-\alpha_1 \sqrt{1-\tau}+\alpha_2 \sqrt{\tau}|^2 \Bigg) \\
\label{PureLossCoh}&& \times \frac{(\alpha_1^*\sqrt{\tau}+\alpha_2^*\sqrt{1-\tau})^n}{\sqrt{n!}} \frac{\alpha_1^m}{\sqrt{m!}} \frac{\alpha_2^l}{\sqrt{l!}}  |m\rangle \langle n|,
\end{eqnarray}
where $d^2 \alpha = d \mathrm{Re}(\alpha) d\mathrm{Im}(\alpha)$. Performing the integrations in Eq.~(\ref{PureLossCoh}), we get a representation of the Kraus operators for the pure loss channel in Fock space,
\begin{eqnarray}
\mathbf{A}_l = \sum_{n=0}^{\infty} \sqrt{\frac{(1-\tau)^l}{l!}} \tau^{\frac{\hat{n}}{2}} \hat{a}^l |n\rangle \langle n |.
\end{eqnarray}
Applying a phase shift $e^{-i \theta x \hat{n}}$ on the environment's mode after interaction, we obtain
\begin{eqnarray}
\mathbf{A'}_l  = \sum_{n=0}^{\infty}  \sqrt{\frac{(1-\tau)^l}{l!}} \tau^{\frac{\hat{n}}{2}} \left(e^{i x \theta}\hat{a}\right)^l |n\rangle \langle n| = e^{i x l \theta} \mathbf{A}_l.
\end{eqnarray}
Applying the same method to the quantum-limited amplification channel we prove that a phase shift $e^{-i \theta y \hat{n}}$ in the environment after interaction leads to,
\begin{eqnarray}
\mathbf{B'}_k &=& \sum_{n=0}^{\infty} \sqrt{ \frac{1}{k!} \frac{1}{G} \left(\frac{G-1}{G}\right)^k} \left( e^{i \theta y} \hat{a}^{\dagger}\right)^k G^{-\frac{\hat{n}}{2}} |n\rangle \langle n| = e^{i y k \theta} \mathbf{B}_k.
\end{eqnarray}
Figure \ref{FigThermalLossesPhases} shows how we decompose the lossy thermal-noise channel and include the phase shift to be estimated. The parameters $x,y\ \in\ \mathbb{R}$, over which we optimise $C_\mathrm{Q}(\theta)$, have the physical meaning of local rotations $x \theta$ ($y \theta$) in the environment of the pure loss (quantum-limited amplification). We close this section by emphasizing the physical meaning of the optimisation. The bound $C_\mathrm{Q}(\theta)$ is the QFI for the state $|\Psi(\theta) \rangle$, which is a purification of $\hat{\rho}$, while $F_\mathrm{Q}(\theta)$ is the QFI for the state $\hat{\rho}$ itself. When we optimise we make the bound tighter, i.e., we look for the smallest possible $C_\mathrm{Q}(\theta)$. This means that we actually look for these phase shift operations on the environment so that maximum information on the estimated parameter is erased from the environmental degrees of freedom, making the QFI $C_\mathrm{Q}(\theta)$ smaller.

\subsection{Calculation of the bound for the single-mode lossy thermal-noise channel}
\label{CalcsSingle}
Performing the derivatives with respect to $\theta$ in equations,
\begin{eqnarray}
\label{H1}\hat{H}_1 &=& \sum_{k,l=0}^{\infty}  \mathbf{A}_l^\dagger \mathbf{B}_k^\dagger \frac{d e^{-i \theta \left( \hat{n} +l x + k y \right)} }{d\theta} \frac{d e^{i \theta \left( \hat{n} +l x + k y \right)}}{d\theta} 
\mathbf{B}_k \mathbf{A}_l,\phantom{12}\ \\
\label{H2}\hat{H}_2 &=& i \sum_{k,l=0}^{\infty}  \mathbf{A}_l^\dagger \mathbf{B}_k^\dagger \frac{d e^{-i \theta \left( \hat{n} +l x + k y \right)}}{d\theta} e^{i \theta \left( \hat{n} +l x + k y \right)} \mathbf{B}_k 
\mathbf{A}_l,\phantom{12}\ 
\end{eqnarray}
using the equations,
\begin{eqnarray}
\label{com1}\hat{a}^l \hat{n}^k= \left(\hat{n} + l \right)^k \hat{a}^l &\Rightarrow& \hat{a}^l G^{\hat{n}}= G^{\hat{n}+l} \hat{a}^l,\\
\label{com2}\left(\hat{a}^\dagger\right)^l \hat{n}^k= \left(\hat{n} - l \right)^k \left(\hat{a}^\dagger\right)^l &\Rightarrow& \left(\hat{a}^\dagger\right)^l \tau^{\hat{n}}= \tau^{\hat{n}-l} \left(\hat{a}^\dagger\right)^l,\\
\label{akadk} \hat{a}^k\hat{a}^{\dagger k} &=& \prod_{j=1}^k (\hat{n}+j),\\
\label{adkak}\hat{a}^{\dagger k} \hat{a}^k &=& \prod_{j=1}^k (\hat{n}-j+1),
\end{eqnarray}
and the key identities of the Kraus operators and their various moments (up to quadratic),
\begin{eqnarray}
\label{sum1} \sum_{l=0}^{\infty} \mathbf{A}_l(\tau)^\dagger  \mathbf{A}_l(\tau) &=& 1,
\end{eqnarray}
\begin{eqnarray}
\label{sum2}\sum_{k=0}^{\infty} \mathbf{B}_k(G)^\dagger \mathbf{B}_k(G) &=& 1,
\end{eqnarray}
\begin{eqnarray}
\label{sum3}\sum_{k,l=0}^{\infty} l \mathbf{A}_l(\tau)^\dagger \mathbf{B}_k(G)^\dagger \mathbf{B}_k(G) \mathbf{A}_l(\tau) &=& \hat{S}_1(\tau)  \equiv \left(1 - \tau\right)\hat{n},
\end{eqnarray}
\begin{eqnarray}
\label{sum4}\sum_{k,l=0}^{\infty} l^2 \mathbf{A}_l(\tau)^\dagger \mathbf{B}_k(G)^\dagger \mathbf{B}_k(G) \mathbf{A}_l(\tau) &=& \hat{S}_2(\tau)  \equiv (1-\tau) \left[ \tau \hat{n}+(1-\tau) \hat{n}^2 \right],
\end{eqnarray}
\begin{eqnarray}
\label{sum8}\sum_{k,l=0}^{\infty} \mathbf{A}_l(\tau)^\dagger \hat{n} \mathbf{B}_k(G)^\dagger \mathbf{B}_k(G) \mathbf{A}_l(\tau) &=& \hat{n} - \hat{S}_1(\tau),
\end{eqnarray}
\begin{eqnarray}
\label{sum9}\sum_{k,l=0}^{\infty} \mathbf{A}_l(\tau)^\dagger \hat{n} \mathbf{B}_k(G)^\dagger \hat{n} \mathbf{B}_k(G) \hat{n} \mathbf{A}_l(\tau) &=& G \left(\hat{n}^2 - 2 \hat{n} \hat{S}_1(\tau) + \hat{S}_2(\tau) \right) +(G-1)\left(\hat{n}-\hat{S}_1(\tau)\right),
\end{eqnarray}
\begin{eqnarray}
\label{sum10}\sum_{k,l=0}^{\infty} k^2 \mathbf{A}_l(\tau)^\dagger \mathbf{B}_k(G)^\dagger \mathbf{B}_k(G) \mathbf{A}_l(\tau) &=& (G-1)^2\left(\hat{n}^2-2 \hat{n}\hat{S}_1(\tau)+\hat{S}_2(\tau)\right) + (G-1)(3G-2)\left(\hat{n}-\hat{S}_1(\tau)\right)\\
\nonumber &&+(G-1)(2G-1),
\end{eqnarray}
\begin{eqnarray}
\label{sum11}\sum_{k,l=0}^{\infty} k \mathbf{A}_l(\tau)^\dagger \mathbf{B}_k(G)^\dagger \mathbf{B}_k(G) \mathbf{A}_l(\tau) &=& (G-1) \left(\hat{n} - \hat{S}_1(\tau) + 1 \right),
\end{eqnarray}
\begin{eqnarray}
\label{sum12}\sum_{k,l=0}^{\infty} \mathbf{A}_l(\tau)^\dagger \mathbf{B}_k(G)^\dagger \hat{n} \mathbf{B}_k(G) \mathbf{A}_l(\tau) &=& G \left(\hat{n} - \hat{S}_1(\tau)+1 \right)-1,
\end{eqnarray}
\begin{eqnarray}
\label{sum13}\sum_{k,l=0}^{\infty} \mathbf{A}_l(\tau)^\dagger \mathbf{B}_k(G)^\dagger \hat{n}^2 \mathbf{B}_k(G) \mathbf{A}_l(\tau) &=& G^2 \left(\hat{n}^2 - 2 \hat{n}\hat{S}_1(\tau)+\hat{S}_2(\tau) \right)+3G(G-1)\left(\hat{n} - \hat{S}_1(\tau)\right) \\
\nonumber && +(G-1)(2G-1),
\end{eqnarray}
\begin{eqnarray}
\label{sum14}\sum_{k,l=0}^{\infty} k \mathbf{A}_l(\tau)^\dagger \mathbf{B}_k(G)^\dagger \hat{n} \mathbf{B}_k(G) \mathbf{A}_l(\tau) &=& G(G-1) \left(\hat{n}^2 - 2 \hat{n}\hat{S}_1(\tau)+\hat{S}_2(\tau) \right)+(G-1)(3G-1)\left(\hat{n} - \hat{S}_1(\tau)\right) \\
\nonumber && +(G-1)(2G-1),
\end{eqnarray}
we find,
\begin{eqnarray}
\label{singleH1} \langle \hat{H}_1 \rangle &=& c_2^2(x,y) \langle \hat{n}^2 \rangle + c_1(x,y) \langle \hat{n}\rangle + c_0(x,y)\\
\label{singleH2} \langle \hat{H}_2 \rangle &=& c_2(x,y) \langle \hat{n} \rangle + d_0(x,y),
\end{eqnarray}
where,
\begin{eqnarray}
\label{c2} c_2(x,y) &=& \frac{\eta +(1-\eta) [(\bar{n}_\mathrm{B}+1) x+\eta  \bar{n}_\mathrm{B}(y+1)]}{(1-\eta ) \bar{n}_\mathrm{B}+1},\\
\label{c1} \nonumber c_1(x,y)&=& \frac{1-\eta }{[(1-\eta ) \bar{n}_\mathrm{B}+1]^2} \Big\{ \eta  (\bar{n}_\mathrm{B}+1) x^2 + \eta  \bar{n}_\mathrm{B} \{1-(1-\eta ) \bar{n}_\mathrm{B} [\eta -4 (1-\eta )\bar{n}_\mathrm{B}-5]\} y^2 \\
\nonumber && + 2 (1-\eta )^2 \bar{n}_\mathrm{B} (\bar{n}_\mathrm{B}+1)^2 x y +2 (\bar{n}_\mathrm{B}+1) [(1-\eta ) \bar{n}_\mathrm{B}+1] [(1-\eta ) \bar{n}_\mathrm{B}-\eta] x  \\ 
          &&  + 2 \eta  \bar{n}_\mathrm{B} [(1-\eta ) \bar{n}_\mathrm{B}+1] [3-\eta +4 (1-\eta )\bar{n}_\mathrm{B}] y + \eta  (4 \bar{n}_\mathrm{B}+1) [(1-\eta ) \bar{n}_\mathrm{B}+1]^2 \Big\}, \\
\label{c0} c_0(x,y) &=& (1-\eta ) \bar{n}_\mathrm{B}  [2 (1-\eta ) \bar{n}_\mathrm{B}+1] (y+1)^2,\\
\label{d0} d_0(x,y) &=& (1-\eta ) \bar{n}_\mathrm{B} (y+1).
\end{eqnarray}
Therefore, the bound can be written as
\begin{eqnarray}
C_\mathrm{Q}(\theta, x,y) = 4 \left[ A(x,y) \langle \Delta n_\mathrm{S}^2 \rangle + \omega(x,y) \right],
\label{singlebound2}
\end{eqnarray}
where,
\begin{eqnarray}
\langle \Delta n_\mathrm{S}^2 \rangle &=& \langle \hat{n}^2 \rangle - \bar{n}_\mathrm{S}^2,\\
\nonumber \label {functionA} A(x,y) &=& \left[\frac{(\bar{n}_\mathrm{B}+1)(1-\eta)}{1+\bar{n}_\mathrm{B} (1-\eta)}\right]^2 x^2 + \left[\frac{\bar{n}_\mathrm{B} \eta (1-\eta)}{1+\bar{n}_\mathrm{B} (1-\eta)}\right]^2 y^2
+ \frac{2\bar{n}_\mathrm{B} (\bar{n}_\mathrm{B}+1) \eta (1-\eta)^2}{[1+\bar{n}_\mathrm{B} (1-\eta)]^2} x y \\
&&  + \frac{2 (\bar{n}_\mathrm{B}+1) \eta (1-\eta)}{1+\bar{n}_\mathrm{B} (1-\eta)} x + \frac{2 \bar{n}_\mathrm{B} \eta^2 (1-\eta)}{1+\bar{n}_\mathrm{B} (1-\eta)} y+\eta^2,\\
\nonumber \omega(x,y)&=& \frac{(\bar{n}_\mathrm{B}+1) \eta (1-\eta) \bar{n}_\mathrm{S}}{[1+\bar{n}_\mathrm{B} (1-\eta) ]^2} x^2 \\
\nonumber &&+ \frac{\bar{n}_\mathrm{B} (1-\eta)\Big\{ n [1+\bar{n}_\mathrm{B} (1-\eta)]^3 + \eta \bar{n}_\mathrm{S} \{ 1 + (1-\eta) \bar{n}_\mathrm{B} [3+ 2 \bar{n}_\mathrm{B} (1-\eta) - \eta] \}  \Big\}}{[1+\bar{n}_\mathrm{B} (1-\eta) ]^2}  y^2\\
\nonumber &&-\frac{2 \eta (1-\eta)^2 \bar{n}_\mathrm{B} (\bar{n}_\mathrm{B}+1) \bar{n}_\mathrm{S}}{[1+\bar{n}_\mathrm{B} (1-\eta) ]^2} x y -\frac{2 \eta (1-\eta) (\bar{n}_\mathrm{B}+1) \bar{n}_\mathrm{S}}{1+\bar{n}_\mathrm{B} (1-\eta)} x\\
\nonumber &&+ \frac{2 \bar{n}_\mathrm{B} (1-\eta) \big\{ [1+\bar{n}_\mathrm{B} (1-\eta)]^2 +\eta \bar{n}_\mathrm{S} [2 + 2 \bar{n}_\mathrm{B} (1-\eta) + \eta] \big\}}{1+\bar{n}_\mathrm{B} (1-\eta) } y\\
&&+ (1-\eta) \left[ \bar{n}_\mathrm{B}^2 (1-\eta) + \eta \bar{n}_\mathrm{S} + \bar{n}_\mathrm{B} (1 + 2 \eta \bar{n}_\mathrm{S}) \right].
\end{eqnarray}
We note that Eqs.~(\ref{sum1}) and (\ref{sum2}) are the completeness relationships for Kraus operators. Equations (\ref{sum3}) and (\ref{sum4}) can be 
proven by choosing a basis to represent the operators (they can also be found in \cite{Escher2011}). Equations (\ref{sum8})-(\ref{sum14}) can be calculated by representing the operators in some basis (e.g. Fock basis) and by performing the summations. Alternatively, in order to prove Eqs.~(\ref{sum8})-(\ref{sum14}), one can use Eqs.~(\ref{sum1})-(\ref{sum4}), the commutation relation $[a,a^\dagger]=1$ and its derivative relation Eq.~(\ref{akadk}).

We now minimise $C_\mathrm{Q}(\theta, x,y)$ by solving for $x,y$ the following equations,
\begin{eqnarray}
\label{minx}\frac{\partial C_\mathrm{Q}(\theta, x,y)}{\partial x} &=& 0, \\
\label{miny}\frac{\partial C_\mathrm{Q}(\theta, x,y)}{\partial y} &=& 0,
\end{eqnarray}
and verifying that the solutions $x_0,y_0$ to Eqs.~(\ref{minx}) and (\ref{miny}) satisfy,
\begin{eqnarray}
\label{test}\frac{\partial^2 C_\mathrm{Q}(\theta, x,y)}{\partial x^2}\Bigg|_{\substack{x=x_0 \\ y=y_0}} \frac{\partial^2 C_\mathrm{Q}(\theta, x,y)}{\partial y^2} \Bigg|_{\substack{x=x_0 \\ y=y_0}} - \left(\frac{\partial^2 C_\mathrm{Q}(\theta, x,y)}{\partial x \partial y}\Bigg|_{\substack{x=x_0 \\ y=y_0}}\right)^2 > 0.
\end{eqnarray}  
We find that
\begin{eqnarray}
\label{x0} x_0 &=& \frac{\eta (\bar{n}_\mathrm{S} - \langle \Delta n_\mathrm{S}^2 \rangle) [\eta \bar{n}_\mathrm{S} + (1-\eta) \bar{n}_\mathrm{B} + 1]}{(1-\eta)\langle \Delta n_\mathrm{S}^2 \rangle [\eta \bar{n}_\mathrm{S} (2 \bar{n}_\mathrm{B}+1) - \eta \bar{n}_\mathrm{B} (\bar{n}_\mathrm{B}+1) + (\bar{n}_\mathrm{B}+1)^2] + \eta \bar{n}_\mathrm{S} [\eta \bar{n}_\mathrm{S} + (1-\eta) \bar{n}_\mathrm{B} + 1]}\\
\label{y0} y_0 &=& -1 - \frac{\eta \langle \Delta n_\mathrm{S}^2 \rangle \bar{n}_\mathrm{S}}{(1-\eta)\langle \Delta n_\mathrm{S}^2 \rangle [\eta \bar{n}_\mathrm{S} (2 \bar{n}_\mathrm{B}+1) - \eta \bar{n}_\mathrm{B} (\bar{n}_\mathrm{B}+1) + (\bar{n}_\mathrm{B}+1)^2] + \eta \bar{n}_\mathrm{S} [\eta \bar{n}_\mathrm{S} + (1-\eta) \bar{n}_\mathrm{B} + 1]}
\end{eqnarray}
and the optimised bound is
\begin{eqnarray}
C_\mathrm{Q}^\star(\theta) = \frac{4 \eta \langle \Delta n_\mathrm{S}^2 \rangle \bar{n}_\mathrm{S} [\eta \bar{n}_\mathrm{S} + (1-\eta)\bar{n}_\mathrm{B} +1]}{(1-\eta)\langle \Delta n_\mathrm{S}^2 \rangle [\eta \bar{n}_\mathrm{S} (2 \bar{n}_\mathrm{B}+1) - \eta \bar{n}_\mathrm{B} (\bar{n}_\mathrm{B}+1) + (\bar{n}_\mathrm{B}+1)^2] + \eta \bar{n}_\mathrm{S} [\eta \bar{n}_\mathrm{S} + (1-\eta) \bar{n}_\mathrm{B} + 1]},
\label{CQstar}
\end{eqnarray}
where $\langle \Delta n_\mathrm{S}^2 \rangle = \langle \hat{n}^2 \rangle - \langle \hat{n} \rangle^2 = \langle \hat{n}^2 \rangle - \bar{n}_\mathrm{S}$ is the variance of the system's photon number.

\subsection{Calculation for the bound for the n-mode lossy thermal-noise channel}
\label{App2}
For the $m$-th lossy thermal-noise channel, using the Eqs.~(\ref{com1}) to (\ref{sum14}), and performing the derivatives in Eqs.~(\ref{H1}) and (\ref{H2}), where now $\hat{H}_1$ and $\hat{H}_2$ correspond to the $m$-th channel (therefore we denote them as $\hat{H}_1^{(m)}$ and $\hat{H}_2^{(m)}$), we have
\begin{eqnarray}
\label{multiH1} \langle \hat{H}^{(m)}_1 \rangle &=& c_2^2(x,y) \langle \hat{n}^2_{(m)} \rangle + c_1(x,y) \langle \hat{n}_{(m)}\rangle + c_0(x,y)\\
\label{multiH2} \langle \hat{H}^{(m)}_2 \rangle &=& c_2(x,y) \langle \hat{n}_{(m)} \rangle + d_0(x,y)\\
\label{multiH2H2} \langle \hat{H}^{(m_i)}_2 \hat{H}^{(m_j)}_2 \rangle &=& c_2^2(x,y) \langle \hat{n}_{(m_i)} \hat{n}_{(m_j)} \rangle + d_0(x,y) d_1(x,y) \left(\langle \hat{n}_{(m_i)} \rangle + \langle \hat{n}_{(m_j)}\rangle \right) + d_0^2(x,y),
\end{eqnarray}
where the functions $c_2(x,y)$, $c_1(x,y)$, $c_0(x,y)$, and $d_0(x,y)$ are given in Eqs.~(\ref{c2}), (\ref{c1}), (\ref{c0}), and (\ref{d0}). The mean values in the right-hand side of Eqs.~(\ref{multiH1}), (\ref{multiH2}), and (\ref{multiH2H2}), are taken on the $m$-th, $m_i$-th, or $m_j$-th mode, as is indicated by their indices. Note that the functions $c_i(x,y)$ and $d_i(x,y)$ depend on $\bar{n}_B$ and $\eta$ as well. From Eqs.~(\ref{multiH1}), (\ref{multiH2}), and (\ref{multiH2H2}), the bound $C_{\mathrm{Q},n}(\theta)$ as defined in the main text, is
\begin{eqnarray}
C_{\mathrm{Q},n}(\theta, x,y) = 4 \left[ A(x,y) \langle \Delta N_\mathrm{S}^2 \rangle + \Omega(x,y) \right],
\label{multibound2}
\end{eqnarray}
where,
\begin{eqnarray}
\label{DNS2}\langle \Delta N_\mathrm{S}^2 \rangle &=& \Big\langle \left( \sum_{m=1}^n \hat{n}_{(m)} \right)^2 \Big\rangle - \langle N_\mathrm{S} \rangle^2,\\
\nonumber \Omega(x,y)&=& \frac{(\bar{n}_\mathrm{B}+1) \eta (1-\eta) \langle N_\mathrm{S} \rangle}{[1+\bar{n}_\mathrm{B} (1-\eta) ]^2} x^2 \\
\nonumber &&+ \frac{\bar{n}_\mathrm{B} (1-\eta)\big\{ n [1+\bar{n}_\mathrm{B} (1-\eta)]^3 + \eta \langle N_\mathrm{S} \rangle \{ 1 + (1-\eta) \bar{n}_\mathrm{B} [3+ 2 \bar{n}_\mathrm{B} (1-\eta) - \eta] \}  \big\}}{[1+\bar{n}_\mathrm{B} (1-\eta) ]^2}  y^2\\
\nonumber &&-\frac{2 \eta (1-\eta)^2 \bar{n}_\mathrm{B} (\bar{n}_\mathrm{B}+1) \langle N_\mathrm{S} \rangle}{[1+\bar{n}_\mathrm{B} (1-\eta) ]^2} x y -\frac{2 \eta (1-\eta) (\bar{n}_\mathrm{B}+1) \langle N_\mathrm{S} \rangle}{1+\bar{n}_\mathrm{B} (1-\eta)} x\\
\nonumber &&+ \frac{2 \bar{n}_\mathrm{B} (1-\eta) \big\{ n [1+\bar{n}_\mathrm{B} (1-\eta)]^2 +\eta \langle N_\mathrm{S} \rangle [2 + 2 \bar{n}_\mathrm{B} (1-\eta) + \eta] \big\}}{1+\bar{n}_\mathrm{B} (1-\eta) } y\\
&&+ (1-\eta) \left[ n \bar{n}_\mathrm{B}^2 (1-\eta) + \eta \langle N_\mathrm{S} \rangle + \bar{n}_\mathrm{B} (n + 2 \eta \langle N_\mathrm{S} \rangle) \right],
\end{eqnarray}
and $A(x,y)$ is given by Eq.~(\ref{functionA}). The mean value in the argument of the summation in Eq.~(\ref{DNS2}) corresponds to the $m$-th input mode, $\langle N_\mathrm{S} \rangle$ and $\langle \Delta N_\mathrm{S}^2 \rangle$ are respectively the total mean photon number and the photon number variance of the $n$-mode input state.
Now we minimise the bound $C_{\mathrm{Q},n}(\theta)$ of Eq.~(\ref{multibound2}) by solving for $x,y$ the equations,
\begin{eqnarray}
\label{minxn}\frac{\partial C_{\mathrm{Q},n}(\theta, x,y)}{\partial x} &=& 0, \\
\label{minyn}\frac{\partial C_{\mathrm{Q},n}(\theta, x,y)}{\partial y} &=& 0.
\end{eqnarray}
The solutions of Eqs.~(\ref{minxn}) and (\ref{minyn}) are
\begin{eqnarray}
x_0 &=& -\frac{\eta^2 \langle N_\mathrm{S} \rangle (\langle \Delta N_\mathrm{S}^2 \rangle-\langle N_\mathrm{S} \rangle)+\eta n (\langle \Delta N_\mathrm{S}^2 \rangle-\langle N_\mathrm{S} \rangle)[1+(1-\eta)\bar{n}_\mathrm{B}]}{D_n},\\
y_0 &=& \frac{\eta^2 \langle N_\mathrm{S} \rangle (\langle \Delta N_\mathrm{S}^2 \rangle-\langle N_\mathrm{S} \rangle) - [n \langle \Delta N_\mathrm{S}^2 \rangle (1+\bar{n}_\mathrm{B}) (1-\eta) + \eta \langle N_\mathrm{S} \rangle (n+2\langle \Delta N_\mathrm{S}^2 \rangle)] [1+\bar{n}_\mathrm{B} (1-\eta)]}{D_n} 
\end{eqnarray}
where,
\begin{eqnarray}
\nonumber D_n&=&\eta^2 \langle N_\mathrm{S} \rangle^2 + \eta n \langle N_\mathrm{S} \rangle [1+(1-\eta)\bar{n}_\mathrm{B}] + (1-\eta)\eta \langle \Delta N_\mathrm{S}^2 \rangle \langle N_\mathrm{S} \rangle (1+2 \bar{n}_\mathrm{B})\\
&& - (1-\eta) \eta \langle \Delta N_\mathrm{S}^2 \rangle n \bar{n}_\mathrm{B} (1+\bar{n}_\mathrm{B})+(1-\eta)n \langle \Delta N_\mathrm{S}^2 \rangle (1+\bar{n}_\mathrm{B})^2.
\end{eqnarray}
The minimised bound $C_{\mathrm{Q},n}(\theta, x_0,y_0) \equiv C_{\mathrm{Q},n}^\star(\theta)$ reads
\begin{eqnarray}
\label{FinalBound}C_{\mathrm{Q},n}^\star(\theta) = \frac{4 n \eta \langle \Delta N_\mathrm{S}^2 \rangle \langle N_\mathrm{S} \rangle [1+\bar{n}_\mathrm{B} (1-\eta)] + 4 \eta^2 \langle \Delta N_\mathrm{S}^2 \rangle \langle N_\mathrm{S} \rangle^2}{D_n},
\end{eqnarray}
where the condition,
\begin{eqnarray}
\label{testn}\frac{\partial^2 C_{\mathrm{Q},n}(\theta, x,y)}{\partial x^2}\Bigg|_{\substack{x=x_0 \\ y=y_0}} \frac{\partial^2 C_{\mathrm{Q},n}(\theta, x,y)}{\partial y^2} \Bigg|_{\substack{x=x_0 \\ y=y_0}} - \left(\frac{\partial^2 C_\mathrm{Q}(\theta, x,y)}{\partial x \partial y}\Bigg|_{\substack{x=x_0 \\ y=y_0}} \right)^2 > 0,
\end{eqnarray}
is satisfied so that $C_{\mathrm{Q},n}^\star(\theta)$ is indeed a minimum.

The bound $C_{\mathrm{Q},n}^\star(\theta)$ is a decreasing function of the environment's mean thermal photon number. Indeed, the derivative of $C_{\mathrm{Q},n}^\star(\theta)$ with respect to $\bar{n}_\mathrm{B}$ reads,
\begin{eqnarray}
\frac{dC_{\mathrm{Q},n}^\star(\theta)}{d\bar{n}_{B}} &=& -\frac{4 \eta (1-\eta) \langle N_\mathrm{S} \rangle \langle \Delta N_\mathrm{S}^2 \rangle^2 }{D_n^2}\Big\{n \eta \langle N_\mathrm{S} \rangle [3+2 \bar{n}_\mathrm{B} (1-\eta)]+2 \eta^2 \langle N_\mathrm{S} \rangle^2 +n^2 [1+\bar{n}_\mathrm{B} (1-\eta)]^2  \Big\}
\end{eqnarray}
which is negative for all input states, $\eta$, $\bar{n}_\mathrm{B}$, and $n$, including of course the single-mode case $n=1$.

\twocolumngrid

\end{document}